\documentclass[conference]{IEEEtran}
\IEEEoverridecommandlockouts
\usepackage{cite}
\usepackage{amsmath,amssymb,amsfonts}
\usepackage{algorithmic}
\usepackage{graphicx}
\usepackage{textcomp}
\usepackage{booktabs}
\usepackage{dblfloatfix} 
\usepackage{xcolor}
\usepackage{url}

\def\BibTeX{{\rm B\kern-.05em{\sc i\kern-.025em b}\kern-.08em
    T\kern-.1667em\lower.7ex\hbox{E}\kern-.125emX}}
\begin{document}

\title{TIPICAL - Type Inference for Python In Critical Accuracy Level\\
}

\author{\IEEEauthorblockN{1\textsuperscript{st} Jonathan Elkobi}
  \IEEEauthorblockA{\textit{GPS} \\
    \textit{UC San Diego}\\
    San Diego, USA \\
    yelkobi@ucsd.edu}
  \and
  \IEEEauthorblockN{2\textsuperscript{nd} Bernd Gruner}
  \IEEEauthorblockA{\textit{Institute of Data Science} \\
    \textit{German Aerospace Center}\\
    Jena, Germany \\
    bernd.gruner@dlr.de}
  \and
  \IEEEauthorblockN{3\textsuperscript{rd} Tim Sonnekalb}
  \IEEEauthorblockA{\textit{Institute of Data Science} \\
    \textit{German Aerospace Center}\\
    Jena, Germany \\
    tim.sonnekalb@dlr.de}
  \and
  \IEEEauthorblockN{4\textsuperscript{th} Clemens-Alexander Brust}
  \IEEEauthorblockA{\textit{Institute of Data Science} \\
    \textit{German Aerospace Center}\\
    Jena, Germany \\
    clemens-alexander.brust@dlr.de}
}
\maketitle

\begin{abstract}
  Type inference methods based on deep learning are becoming increasingly popular as they aim to compensate for the drawbacks of static and dynamic analysis approaches, such as high uncertainty. However, their practical application is still debatable due to several intrinsic issues such as code from different software domains will involve data types that are unknown to the type inference system.

  In order to overcome these problems and gain high-confidence predictions, we thus present TIPICAL, a method that combines deep similarity learning with novelty detection. We show that our method can better predict data types in high confidence by successfully filtering out unknown and inaccurate predicted data types and achieving higher F1 scores to the state-of-the-art type inference method Type4Py. Additionally, we investigate how different software domains and data type frequencies may affect the results of our method.
\end{abstract}

\begin{IEEEkeywords}
  type inference, novelty detection, machine learning, cross-domain
\end{IEEEkeywords}

\section{Introduction}
Dynamic programming languages can be enriched by optional type annotations to enable more precise program analysis and early detection of type-related run-time errors \cite{Hanenberg2013AnES,TypeNotType}. Our objective is to develop a workable technique for Python programmers to use on a daily basis that will enhance their routine workflow through the annotation of optional data types, a process known as type inference. Giving type recommendations to the user in real-time as well as automatically after writing the code will accomplish this. Due to its automation, the first case, however, necessitates a method that only annotates with high certainty of correctness, as the harm that inaccurate type hints can do may be greater than the benefit of the positive one.

Static and dynamic type inference techniques suffer from low precision due to applied abstraction or missing coverage~\cite{10.1145/2989225.2989226}. Recent deep learning-based methods aim to overcome these issues and provide promising results \cite{Typilus,Opttyper,Type4py,TypeBert}. However, these systems struggle with problems occurring in practical applications for example data types unknown to the system or source code from other software domains \cite{gruner2022cross}.

First, the problem of unknown classes, or the inability to accurately predict unseen data types, is a prevalent issue in the field of machine learning. This is due to the lack of representation of such data types in the training set, rendering prediction of these data types ineffective. To mitigate this issue, we propose a method to filter out unknown data types based on their characteristic features. This approach not only enables the identification of unknown data types but also improves the overall reliability and quality of data type annotations by eliminating inaccurate predictions.

In practical usage, type inference systems are often utilized in a variety of software domains. However, this can exacerbate the problem of unknown data types and lead to decreased accuracy of predictions due to dataset shifts \cite{dataset_shifts_diagnostics,finlayson2021clinician}. Therefore, in our research, we investigate the effect of different software domains on the performance of our type inference method.

We thus present \textbf{TIPICAL} - Type Inference for Python In Critical Accuracy Level, an extension of the Type4Py method \cite{Type4py} for obtaining accurate results that address both the unknown data types problem and the domain shift in order to maximize the practical application of deep learning-based type inference. In our experiments, we show that our method can successfully filter out unknown and inaccurate predicted data types and improve the results compared to the state-of-the-art type inference method Type4Py.

Furthermore, we investigate the impact of the data type frequencies and different software domains. For the evaluation, we use the recent datasets CrossDomainTypes4Py \cite{gruner2022cross} and ManyTypes4Py \cite{mt4py2021}. In order to ensure the \textbf{reproducibility} of our experiments, we make our experimental pipeline publicly available\footnote{\url{https://gitlab.com/dlr-dw/type-inference}}.

\section{Related Work}
\label{sec:related Work}
\begin{figure*}[t]
  \centering
  \includegraphics[width=0.7\textwidth]{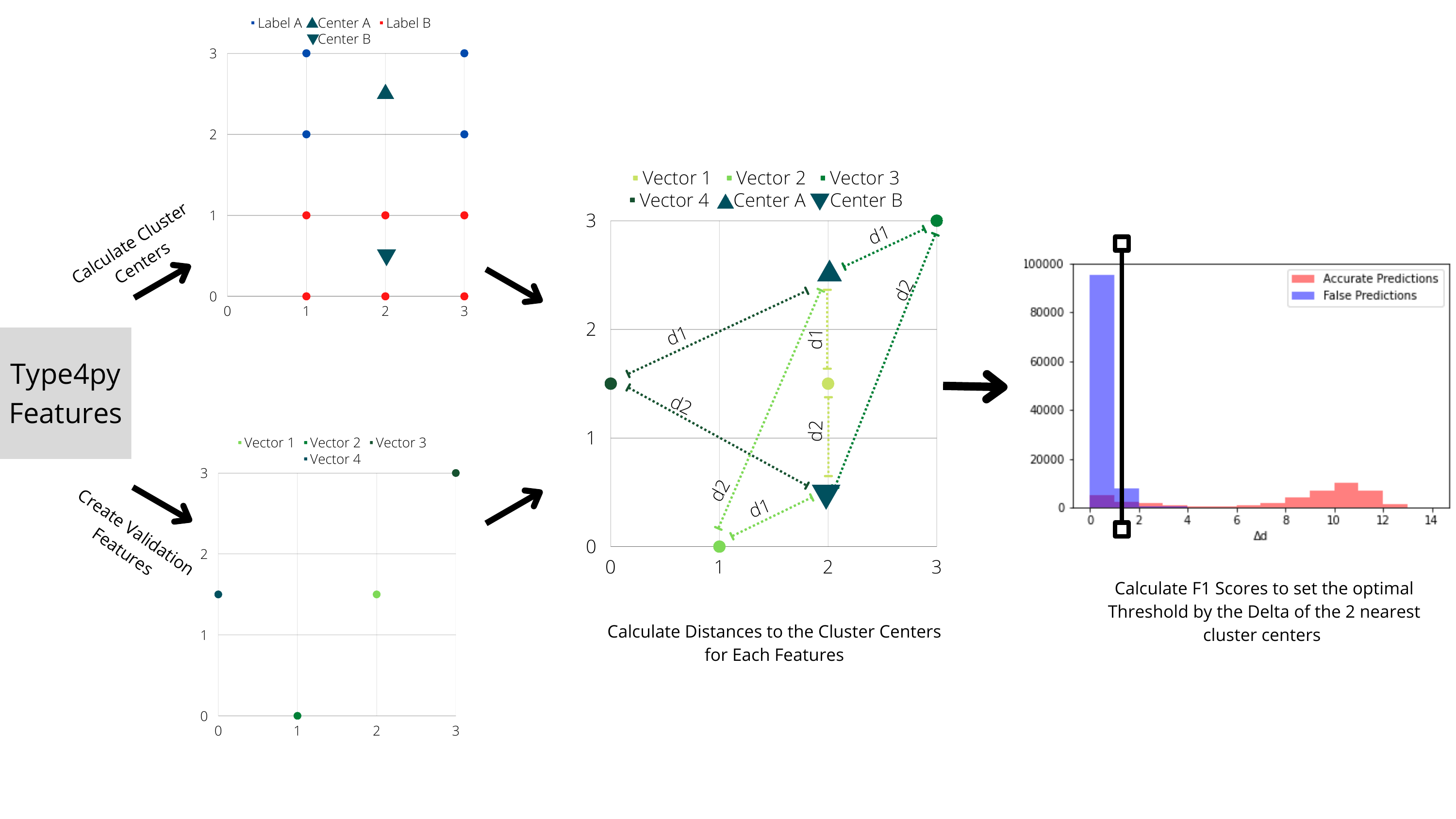}
  \caption{Deciding the threshold using the validation set
  }
  \label{fig:decide threshold}
\end{figure*}

Several studies use deep learning techniques for type inference. In this work, we focus on methods that are designed for Python projects. DLType \cite{Dltpy}, TypeWriter \cite{TypeWriter}, and PyInfer \cite{Pyinfer} are the first deep learning-based methods in that field. They suffer from the problem that they can only predict a limited number of data types due to their architecture. These methods are limited to the 500 or 1000 most frequent data types that occur in the training dataset.

Typilus \cite{Typilus}, and Type4Py \cite{Type4py} address this problem and can predict all data types which are present in the training dataset. There are potential approaches where new types can be recognized through additional static analysis, as demonstrated in HiTyper \cite{Hityper}. However, there is still the issue that data types that are rarely or not in the training dataset cannot be predicted\cite{gruner2022cross}.

Nevertheless, Novelty detection has not been actively pursued as a solution to the problem. Therefore we develop TIPICAL to mitigate these problems. As a basis for our method, we use Type4Py, because according to the evaluation of Mir et al. \cite{Type4py}, it is state-of-the-art and the source code is publicly available.

\section{Methodology}
\label{sec:methods}

In this section, we briefly explain the type inference method Type4Py, which is the basis for our method. Afterward, the structure and the functioning of TIPICAL are presented.

\subsection{Type4Py Inference System}

The Type4Py framework serves as the foundation for our research, as detailed in the original paper \cite{Type4py}. The Type4Py system utilizes code tokens, identifier names, and available data types (visible type hints) as input. These code tokens and identifier names are embedded by Word2Vec \cite{Word2Vec} and processed separately through recurrent neural networks. Then the resulting representations are concatenated with the visible type hints and further processed through a fully connected layer to generate a feature vector. This feature vector is then used for a k-nearest neighbor search in the type cluster to predict all data types using the training dataset. However, it should be noted that this method may not accurately predict unknown data types that are not represented in the training dataset, as well as those that are inaccurately predicted due to limitations of using the nearest neighbor as a classification method

\begin{figure*}[t!]
  \centering
  \includegraphics[width=0.7\textwidth]{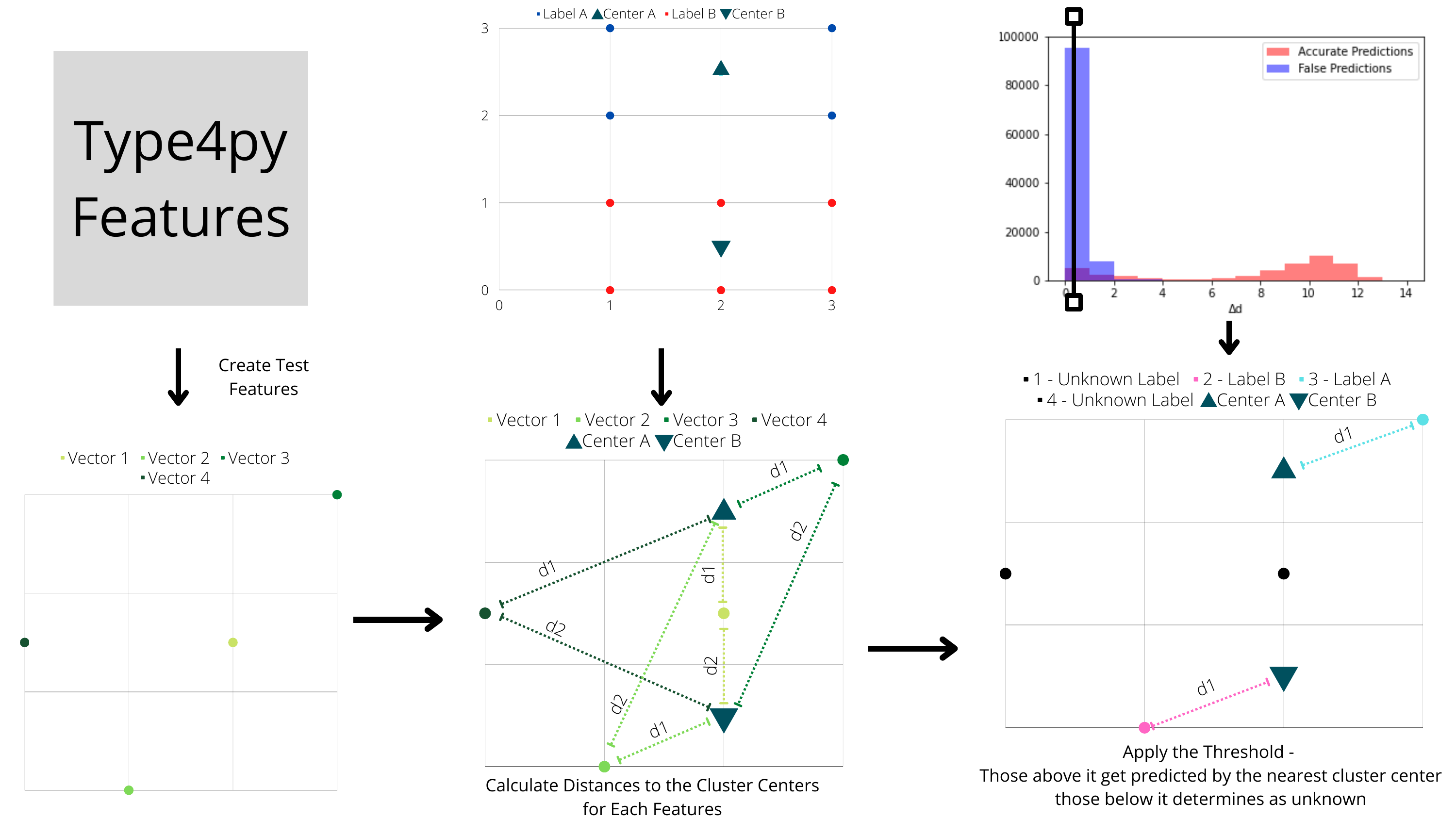}
  \caption{Applying the threshold to make predictions on the test set}

  \label{fig:apply threshold}
\end{figure*}

\subsection{TIPICAL}
\label{subsec:tipical}
As a result, we sought to develop TIPICAL, a comprehensive system that would enhance the usage of type inference by utilizing novelty detection to filter out inaccurate predictions and unpredictable data types. Using the workflow presented in Figure \ref{fig:decide threshold}, we find a threshold for filtering, We then use the same cluster centers threshold on the test set to filter out predictions.

\subsubsection{Determining the Threshold}
First, we determine the cluster centers of each known data type using the training dataset, defined as follows:
\begin{equation}
  \bar{\vec{x}} = \frac{\sum_{i=1}^{n}\vec{x_{i}}}{n},
\end{equation}  \newline
where $\vec{x}$ is the feature vector, $n$ is the total number of vectors and $\bar{\vec{x}}$ is the cluster center vector. Next, we determine the top two nearest cluster centers for each vector in the validation set, and their distances $d1$ and $d2$.
\begin{equation}
  \label{equ:delta}
  \Delta d = d_{2}-d_{1}
\end{equation}
Then $\Delta d$ is calculated, as seen in Equation \ref{equ:delta}. It can be interpreted as cost-effective active learning or the proxy of the entropy of the distribution of the distances between all of the cluster centers.

After which, using these facts, we develop a threshold-based method to determine whether or not the closest cluster center can accurately predict the data type. The reasoning behind this is that if two cluster centers are roughly the same distance from the vector, it may be difficult to distinguish which datatype is the correct one or even they could be an unknown data type that is just not represented in the training set, whereas if the closest cluster center is significantly closer to the vector than the second one, it will almost certainly make the correct prediction. The outliers are labeled as non-predicted for further research directions as can be seen later in the conclusion.

\subsubsection{Making the Predictions}
Afterward, for the prediction itself, we use the nearest cluster center as the predicted data type. In order to make accurate predictions, we maximize the F1 score on the validation set by determining until which value of $\Delta d$ we are going to filter out the vectors. Finally, as can be seen in the second part of Figure \ref{fig:decide threshold}, we apply the same approach to the test set using our previous findings. We calculate $d1$, $d2$, and  $\Delta d$ for all the test set vectors. Further, we apply the threshold that we determined from the training set and filter out lower  $\Delta d$ from our final predictions, producing only high-certainty predictions that will provide our end-user with the most accurate predictions, as will be demonstrated in the following section.

\section{Experiments and Evaluation}
\label{sec:experiments}
Following the pipeline of TIPICAL described in Section \ref{subsec:tipical}, we present our experiments on the ManyTypes4Py and the CrossDomainTypes4Py datasets. Moreover, we expanded the scope of the experiments of the original papers\cite{Type4py,gruner2022cross} to further study the effects of different software domains, and the unknown data types issues.

In addition to creating a comprehensive method for real-life machine learning-based type inference with high certainty, we conducted the experiments to answer the following research questions:

\begin{enumerate}
  \item  How do different software domains affect the predictable and unknown data type distribution according to the entropy proxy - $\Delta d$?
  \item  How does the nearest cluster center-based method accuracy of the data types predictions correspond to $\Delta d$?
  \item  Can TIPICAL create higher certainty predictions than the predictions of the Type4Py system?
  \item  How does the frequency of data types affect its predictability in TIPICAL?
\end{enumerate}

\subsection{Datasets and Domains}
We use the CrossDomainTypes4Py \cite{gruner2022cross} and ManyTypes4Py \cite{mt4py2021} datasets for our experiments. As described by Gruner et al. \cite{gruner2022cross}, these contain a total of at least three different software domains. CrossDomainTypes4Py consists of the scientific calculation (cal) domain with 4,783 repositories and the web development (web) domain with 3,129. ManyTypes4Py, on the other hand, contains 5,382 repositories, which are from various domains and are therefore considered general (mtp).

\subsection{Experiment Setup}
We adapt the existing cross-domain Type4Py implementation from Gruner et al. \cite{gruner2022cross} in order to conduct the research. We employ PyTorch, a deep learning framework, using Python 3.6. We use the same hyperparameters as Mir et al \cite{Type4py}.

For our experiments, we created the following four cross-domain setups:
\begin{enumerate}
  \item Setup Cal2Mtp - Scientific Calculation to General
  \item  Setup Mtp2Cal - General to Scientific Calculation
  \item  Setup Cal2Web - Scientific Calculation to Web Development
  \item  Setup Web2Cal - Web Development to Scientific Calculation
\end{enumerate}

\subsection{Research Questions and Results}
\textbf{RQ1: How do different software domains affect the predictable and unknown data types distribution according to the entropy proxy - $\Delta d$?}\\\
Each setup consists of two experiments, which are conducted three times to calculate the average of the results. In the first experiment, the system is trained on the first mentioned domain and evaluated on the second domain (example reference: Cal2Mtp.1). The second experiment is for comparison and performs the training and the evaluation on the second mentioned domain (example reference: Cal2Mtp.2). In total we get eight results from our four setups.

\begin{figure}[t]
  \centering
  \includegraphics[width=0.7\linewidth]{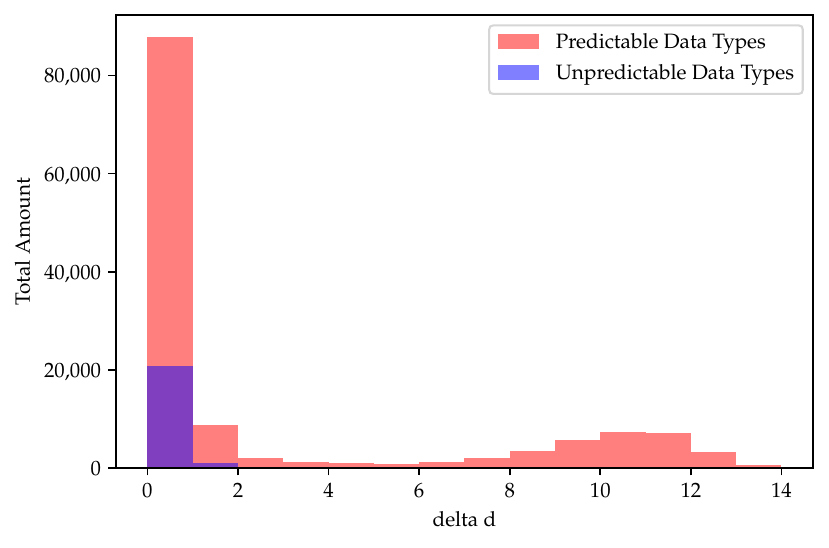}
  \caption{Histogram of precitable vs. unknown data types distribution according to $\Delta d$ for setup Cal2Mtp.1
  }
  \label{fig3}
\end{figure}

\begin{figure}[t]
  \centering
  \includegraphics[width=0.7\linewidth]{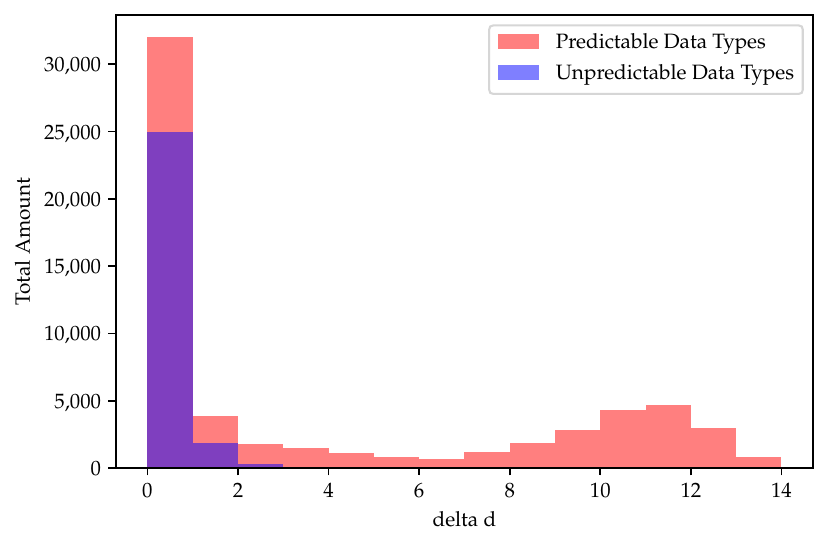}
  \caption{Histogram of precitables VS unknown data types distribution according to $\Delta d$ for Setup Cal2Mtp.2
  }
  \label{fig4}
\end{figure}

\begin{table*}[t]
  \centering
  \caption{Method Scores Comparison}
  \begin{tabular}{lllllllll}
    \hline
    \textbf{Methods/Setups}            & Cal2Mtp.1 & Cal2Mtp.2 & Mtp2Cal.1 & Mtp2Cal.2 & Cal2Web.1 & Cal2Web.2 & Web2Cal.1 & Web2Cal.2 \\ \hline
    TIPICAL {[}F1{]}                   & 67.09\%   & 88.05\%   & 87.41\%   & 64.26\%   & 88.65\%   & 77.45\%   & 76.77\%   & 86.17\%   \\
    Type4Py {[}F1{]}                   & 43.96\%   & 44.69\%   & 45.27\%   & 58.74\%   & 47.41\%   & 49.33\%   & 47.20\%   & 55.35\%   \\
    Type4Py only predictables {[}F1{]} & 65.26\%   & 64.55\%   & 63.27\%   & 73.71\%   & 67.18\%   & 67.17\%   & 65.09\%   & 71.47\%   \\ \hline
  \end{tabular}
\end{table*}

Figures 3 \& 4 show that using examples from various software domains has no discernible impact on the target domains. The amount of unpredictable types, which sharply increases due to the size of the source dataset as a whole rather than a change in the use case itself, is another intriguing development that follows from this. Nevertheless, those findings are in favor of our approach since, by removing the lower scores of the $\Delta d$, we will also remove most of the unknown data types since they are closer to at least two cluster centers within a comparable distance. Hence we provide a novelty detection method for practical use.

\textbf{RQ2: How does a nearest cluster center-based method accuracy of the data types predictions corresponds to $\Delta d$?}
\begin{figure}[t]
  \centering
  \includegraphics[width=0.7\linewidth]{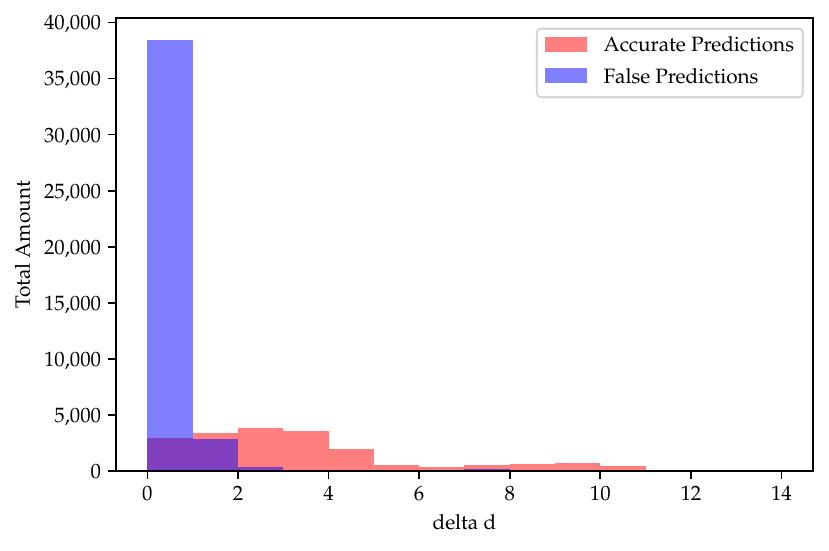}
  \caption{Histogram of accurate and inaccurate predictions distribution according to $\Delta d$ for Setup Web2Cal.1
  }
  \label{fig5}
\end{figure}
\begin{figure}[t]
  \centering
  \includegraphics[width=0.7\linewidth]{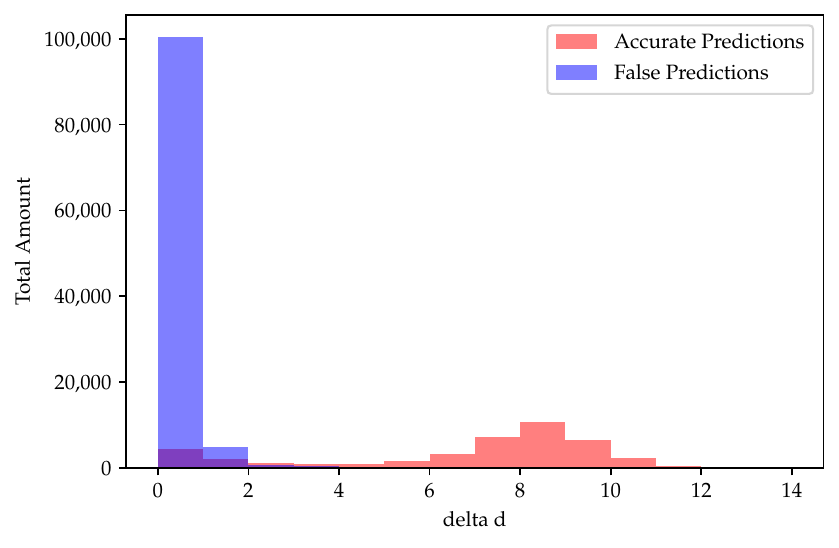}
  \caption{Histogram of accurate and inaccurate predictions distribution according to $\Delta d$for Setup Web2Cal.2
  }
  \label{fig6}
\end{figure}

Figures 5 \& 6 illustrate that various domains do not affect the accuracy of the target domains of the closest cluster center's prediction. However, the closest cluster center for the right predictions examples causes the distribution of $\Delta d$ to be shifted to a greater distinction. As a result, we can conclude that a domain change use case might be predicted with a better degree of certainty using our method. In addition, those results support our method because by filtering the lower scores of $\Delta d$. We will also eliminate the majority of incorrect predictions since they are located nearer to at least two cluster centers than the accurate ones.

\begin{figure}[t]
  \centering
  \includegraphics[width=0.7\linewidth]{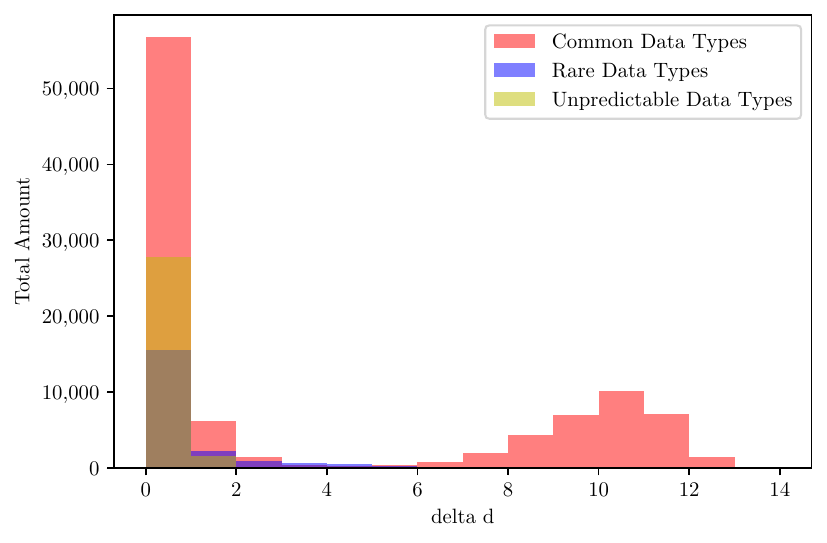}
  \caption{Histogram of common and rare types distributed across the subtraction of the 2 Nearest cluster centers for Setup 3.1
  }
  \label{fig7}
\end{figure}

\textbf{RQ3: Can TIPICAL create higher certainty predictions than the predictions of the Type4Py system?}

\begin{table*}[t]
  \centering
  \caption{Example Categories Distribution}
  \begin{tabular}{lllllllll}
    \hline
    \textbf{Samples/Setups}                      & Cal2Mtp.1 & Cal2Mtp.2 & Mtp2Cal.1 & Mtp2Cal.2 & Cal2Web.1 & Cal2Web.2 & Web2Cal.1 & Web2Cal.2 \\ \hline
    \textbf{Predicted Samples}                   & 38.31\%   & 30.62\%   & 32.48\%   & 41.83\%   & 30.09\%   & 37.10\%   & 35.10\%   & 27.50\%   \\
    Samples Predicted Accurate                   & 24.62\%   & 26.96\%   & 28.77\%   & 27.67\%   & 25.93\%   & 28.73\%   & 26.96\%   & 24.67\%   \\
    Predictable Samples Predicted Inaccurate     & 11.55\%   & 2.62\%    & 2.18\%    & 11.60\%   & 3.39\%    & 5.39\%    & 5.84\%    & 2.16\%    \\
    Unknown Samples Predicted                    & 2.14\%    & 1.04\%    & 1.53\%    & 2.56\%    & 0.77\%    & 2.98\%    & 2.30\%    & 0.66\%    \\ \hline
    \textbf{Non-Predicted Examples}              & 61.69\%   & 69.38\%   & 67.52\%   & 58.17\%   & 69.91\%   & 62.90\%   & 64.90\%   & 72.50\%   \\
    Samples Non-Predicted Accurate               & 2.96\%    & 3.12\%    & 2.82\%    & 2.95\%    & 3.67\%    & 5.08\%    & 4.26\%    & 3.39\%    \\
    Predictable Samples Non-Predicted Inaccurate & 46.52\%   & 36.26\%   & 35.19\%   & 43.42\%   & 47.04\%   & 31.29\%   & 33.43\%   & 49.80\%   \\
    Unknown Samples Non-Predicted                & 12.22\%   & 30.00\%   & 29.52\%   & 11.80\%   & 19.20\%   & 26.53\%   & 27.21\%   & 19.31\%   \\ \hline
  \end{tabular}
\end{table*}
Table 1 presents a comparison of the final results of our proposed method, TIPICAL, with the benchmark method. It is evident from this table that TIPICAL produces significantly better high-certainty type annotations for the end user, even across different software domains. On the other hand, the benchmark method exhibits a significant amount of noise, resulting from numerous inaccurate predictions of both predictable and unpredictable data types. To obtain the final F1 scores, our method employs a filtering approach, as described earlier, by utilizing a threshold determined from the validation group of the training dataset, effectively reducing the majority of the noise. Finally, TIPICAL reach better results in all of the experiments, with an average of 79.48\% F1 score, which is 30.49\% improvement from the average of the Type4Py method.

To further demonstrate the effectiveness of our proposed method, TIPICAL, in comparison to the Types4Py technique, Table 1 presents a comparison of the results obtained when only predicting the predictable data types in the test dataset. Even in this scenario where the predictable data types are known, TIPICAL outperforms Types4Py in 7 out of 8 experiments. On average, our method gets 12.27\% F1 score. This serves as evidence that TIPICAL, which does not make any assumptions about the target dataset, is more effective than methods that rely on prior knowledge of the predictable data types.

\textbf{RQ4: How does the frequency of data types affect its predictability in TIPICAL?}

The frequency of a data type influences how accurately predictable it is. Figures 7 and 8 show that common types (types that appear in the training more than 100 times \cite{Type4py}) are presented after the threshold at a higher rate than rare or unknown types. Additionally, although the rare type has a longer tail, we can see similarities between the distribution of the unknown and rare types. Because the calculated cluster centers of the common data types more closely resemble the novel label features, we can also predict the common data types more accurately than the rare types. Furthermore, the distribution of the common type is impacted by different software domains, whereas the distribution of the rare type is not. This may help to explain why some cases have decreased accuracy.

\section{Limitations}
\label{sec:limitations}

\begin{figure}[t]
  \centering
  \includegraphics[width=0.7\linewidth]{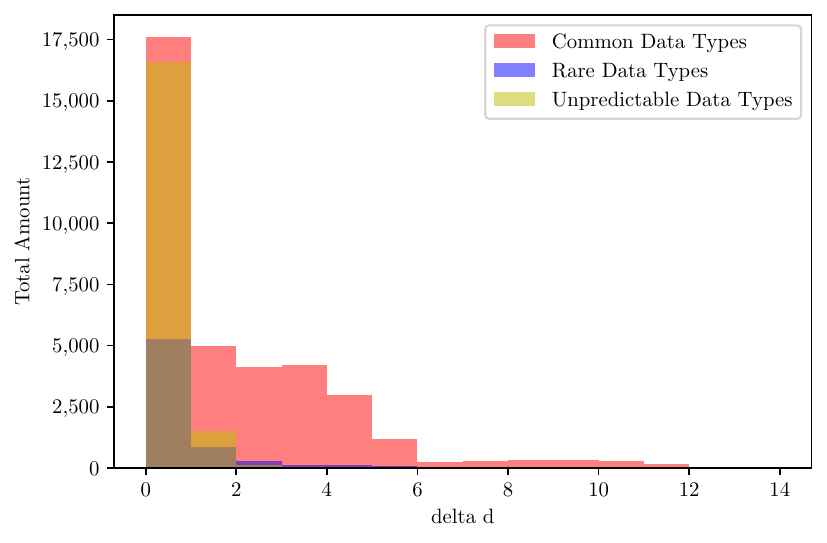}
  \caption{Histogram of common and rare type distributions across the subtraction of the 2 Nearest cluster centers for Setup 3.2
  }
  \label{fig8}
\end{figure}

In order to perform a comparative analysis with the current benchmark of Type4Py, we conducted a repetition of previous experiments while disregarding the validation data type. This enabled us to make a comparison with the methodology employed by Type4Py.

As demonstrated in Table 2, our method predicts approximately one-third of the examples within the dataset. Although, as indicated by the "Samples Non-Predicted Accurate" row, a trade-off of 3.53\% of accurate predictions must be made on average to apply the threshold as TIPICAL is designed to predict only a portion of the information. Nonetheless, by excluding, on average, 92.63\% of the unknown data types (21.97\% of the dataset) and 40.37\% of potentially inaccurate predictions, as indicated by the "Unknown Samples Non-Predicted" and "Predictable Samples Non-Predicted Innacurate" rows, respectively, we observed an increase in the confidence of our predictions by 30.49\% (F1) in comparison to the Type4Py method. Our findings were consistent with this conclusion, as exemplified by setup 1.2 where TIPICAL demonstrated a 43.36\% improvement in performance while only predicting 30.62\% of the examples. As a result, our research confirms that a lower prediction count results in higher scores.

\section{Conclusion}
\label{sec:conclusion}
This study offers a thorough procedure for carrying out the most precise deep learning-based type inference, TIPICAL. It reaches an average 79.48\% F1 score, which is 30.49\% improvement from the Types4Py method. We accomplish this by adding steps to the Type4Py pipeline's final stages that address the two main issues with the current approaches: predictability and uncertainty.
TIPICAL is a novel approach to type inference in Python, utilizing a combination of machine learning and novelty detection techniques to improve the accuracy of predictions. The system is built upon the Type4Py framework, which uses the closest neighbor as a sole method for classification, which can lead to inaccuracies and unpredictable data types. To mitigate this issue, TIPICAL employs a threshold-based method to determine the accuracy of predictions by comparing the distance to the closest cluster center to the second closest. This approach results in the filtering out of lower certainty predictions, thus maximizing the overall accuracy of predictions on both the validation and test sets.
Consequently, the best current method for use in practice that needs high certainty of data type annotation is produced. Nevertheless, with our 8 experiments across 3 domains, we focus on 3 subjects. Real-world software domain changes, which mainly affect the accuracy of the predictable data types but not the prediction percentage of the unpredictable data types, lead to the conclusion that our technique can compete well even then. Additionally, the fact that the main types are common and easier to predict, further increases the validity of our method, mainly due to the robustness of predicting well-represented types in the training set.

Our findings suggest that applying the type inference automatically using TIPICAL would be preferable because it will cause less harm, however, it may not be as effective for real-time suggestions when the developer may review all of the type annotations.

Last but not least, we can suggest further developing this method by utilizing lifelong learning techniques to improve the general predictability of labels and enrich the predictable types. Moreover, another easy to achieve better results is to use only the common types from the training set to create the cluster centers, hence, the sureness of the cluster center will raise, and the approach will predict probably higher F1 scores, but with fewer samples. Due to the countless possible types, for example, by providing an example that is below the threshold so that a domain expert can label it and repeat the entire process indefinitely\cite{Kaeding2016}. Researchers can also use a broad method to determine the genuine entropy, then follow the instructions in our method to increase the method's effectiveness and further study it.

\bibliographystyle{IEEEtran}
\bibliography{IEEEabrv, type_inference}
\end{document}